\documentclass[aps,preprint,showpacs,superscriptaddress]{revtex4}
\usepackage{amssymb,bm}
\usepackage{graphicx}
\usepackage{amsmath}
\usepackage{comment}
\allowdisplaybreaks
\usepackage[utf8]{inputenc}

\usepackage{color}

\begin{document}

\title{Gauge invariance and amplitudes of two-photon processes}
\author{P. A. Krachkov}
\email{P.A.Krachkov@inp.nsk.su}
\affiliation{Budker Institute of Nuclear Physics of SB RAS, 630090 Novosibirsk, Russia}
\author{A. I. Milstein}
\email{A.I.Milstein@inp.nsk.su}
\affiliation{Budker Institute of Nuclear Physics of SB RAS, 630090 Novosibirsk, Russia}
  \author{A. G. Shamov}
\email{A.G.Shamov@inp.nsk.su}
\affiliation{Budker Institute of Nuclear Physics of SB RAS, 630090 Novosibirsk, Russia}

\date{\today}

\begin{abstract}
A method to derive the convenient representations for many  two-photon amplitudes is suggested. It is based on the use of the gauge in which  the photon propagator   has only space components.  The amplitudes obtained   have no any strong numerical cancellations and, therefore,  are very convenient for numerical evaluations. Our approach  is illustrated by the consideration of the processes  $e^+e^-\to e^+e^-e^+e^-$, $e^+e^-\rightarrow e^+e^-\mu^+\mu^-$, and $e^+e^-\to \mu^+\mu^-\pi^+\pi^-$. The method is  extended on the case of polarized particles. The amplitudes obtained in this approach have been employed for extension
of the event generator developed by F.A.~Berends, P.H.~Daverveldt, and R.~Kleiss.
\end{abstract}


\maketitle

\section{Introduction}
Investigations of  processes $e^+e^-\rightarrow e^+e^-X$  (two-photon processes),  
where $X$ are pairs of leptons ($e,\,\mu,\, \tau$)  or hadrons with C-parity   $C = +1$ give an important
experimental information on physics of $\gamma \gamma$ collisions. Such investigations are  the important parts of the  physical programs of C- and B- factories.
Since the leptonic two-photon processes are completely described by QED, they   provide  firstly  a possibility to test  QED in higher orders of the perturbation theory  and secondly are important  for calibration of the experiments and suppression  of backgrounds.

There are a few generators that simulate leptonic two-photon processes beyond the equivalent photon approximation \cite{BDK86,Schu98,Ver83,HKD93,BPP2001}. The most popular   is a  generator of F.A.~Berends, P.H.~Daverveldt, and R.~Kleiss (DIAG36 or BDK generator ) \cite{BDK86}. It includes   all Feynman diagrams and accounts for identity effect. It is worth noting that this effect  is not taken into account in other two-photon event generators. 

The BDK prediction of the identity effect in $e^+e^-\rightarrow 2(e^+e^-)$
  process has been presented in Ref.~\cite{KMRS2019}. The unexpected growth of the effect with
the cut on the invariant mass of produced $e^+e^-$ pair motivated us to make calculations with
the BDK generator in which the alternative matrix elements expression were embedded.
The BDK prediction on the identity effect were confirmed at the increased accuracy of the
numerical calculations.

These alternative matrix element expressions are presented below. Unlike to those used
in BDK \cite{BDK85}, they contain only spatial components of the involved particle momenta which
ensure their numerical stability. Another advantage of them is an easiness of the modification for semi-leptonic final states. Simulation of the semi-leptonic processes such as  $ e^+ e^- \rightarrow\mu^+ \mu^- \pi^+ \pi^- $ became actual in context of BELLE analysis of $ e^+ e^- \rightarrow \Upsilon (nS)\pi^+ \pi^-$ with $\Upsilon (nS)\rightarrow l^+l^-$ \cite{Mizuk2019}.
 In this case two-photon  processes are main sources of the background.

\section{Method of calculation}

First of all, to obtain the explicit form of the matrix element, it is necessary to introduce some convenient basis for the Dirac spinors. Since the number of the Feynman diagrams is quite large, for numerical calculations it is important to have the most simple expression for the total amplitude.  Then, the following circumstance must be taken into account.  In the two-photon processes, the main contribution to  the cross section is given by kinematics where a virtual photon is emitted almost along the initial electron momentum $\bm p$. In such kinematics the corresponding   current $J_\mu$ has  large  components $J_0$ and $J_\parallel$, i.e.,  $J_0\sim J_\parallel\sim \gamma  J_\perp \sim \gamma^2\,(J_0-J_\parallel )$, where $J_\parallel=\bm J\cdot\bm p/p$ and $\gamma=\varepsilon/m_e$ ($m_e$ is the electron mass, we set $\hbar=c=1$).  As a result, using the   Feynman gauge leads to a    huge numerical cancellation between $J_0^2$ and  $J_\parallel^2$ at high energies, which is a problem for numerical calculations with high accuracy. In our work  we avoid this  problem  by means of the gauge in which the photon propagator $D_{\mu\nu}(k)$ contains only spatial components:
\begin{align}\label{prop}
D_{ij}(k)=-\frac{1}{\omega^2-\bm k^2+i\,0}\left(\delta_{ij}-\frac{k_i k_j}{\omega^2}\right)\,,\quad D_{0\mu}(k)=0\,,\quad k^\mu=\,(\omega,\,\bm k)\,.
\end{align}
To calculate the amplitude of the process, we use the following explicit form of the positive-energy Dirac spinor   $U_{a, \lambda_a}$  and negative-energy Dirac spinor  $V$ \cite{LL}
\begin{eqnarray}\label{uv}
U_{a,\, \lambda_a} =N_a 
\begin{pmatrix}
\phi_{\lambda_a}\\
\bm \sigma\cdot\bm P_a \, \phi_{\lambda_a }\end{pmatrix}\, ,\quad
V_{b,\, \lambda_b}=N_b
\begin{pmatrix}
\bm \sigma\cdot\bm P_b\,\phi_{\lambda_b}\\
\phi_{\lambda_a}\end{pmatrix}\, ,\nonumber\\
\bm P_a=\dfrac{\bm p_a}{\varepsilon_{a}+m}\,,\quad \bm P_b=\dfrac{\bm p_b}{\varepsilon_{b}+m}\,,\quad 
N_a=\sqrt{\dfrac{\varepsilon_{a}+m}{2\varepsilon_{a}}}\,,\quad N_b=\sqrt{\dfrac{\varepsilon_{b}+m}{2\varepsilon_{b}}}\,.
\end{eqnarray} 
Here $\bm{\sigma}$ are Pauli matrices, $\bm p_a$ and $\varepsilon_{a}$ are the electron momentum and energy, $\bm p_b$ and $\varepsilon_{b}$ are the positron momentum and energy, $\phi_{\lambda }$ is the two-component spinor, $\lambda=\pm 1$ denotes two possible polarizations,   the same basis of spinors for all particles is used. We introduce three unit orthogonal vectors $\bm e_x$, $\bm e_y$, $\bm e_z$ (so that $[\bm e_x\times\bm e_y]=\bm e_z$) and choose $\phi_{\lambda }$ to be the eigenstate of the operator $\bm \sigma\cdot\bm e_z$, i.e.,  $(\bm \sigma\cdot\bm e_z)\,\phi_{\lambda }=\lambda\,\phi_{\lambda }$. It is convenient to direct $\bm e_z$ along the momentum of initial electron. Then we have
\begin{eqnarray}\label{eq1}
&\phi_{\lambda_a}\phi_{\lambda_b}^+=\dfrac{1}{2}(A_{ab}+\bm\sigma\cdot\bm B_{ab})\,,\nonumber\\
&A_{ab}=\delta_{\lambda_a\lambda_b}\,,\quad  \bm B_{ab}=\lambda_a\delta_{\lambda_a\lambda_b}\bm e_z+
\delta_{\lambda_a\bar\lambda_b}(\bm e_x+i\lambda_a\bm e_y)\,,\quad \bar{\lambda}=-\lambda\,, 
\end{eqnarray}
Using the quantities $A_{ab}$ and $\bm B_{ab}$, we obtain the matrices $U_a\bar U_b\,,U_a\bar V_b\,,V_a\bar U_b\,,V_a\bar V_b\,,$ which   are necessary for further calculations:
\begin{align}\label{eq2}
&U_a\bar U_b=\dfrac{1}{4}N_a N_b[\gamma^0f^{(0)}_{ab}-f^{(1)}_{ab}+\gamma^0\gamma^5f^{(2)}_{ab}+\gamma^5f^{(3)}_{ab}\nonumber\\
&+\gamma^0\bm
\Sigma\cdot\bm g^{(0)}_{ab}-\bm\Sigma\cdot\bm g^{(1)}_{ab}-\bm{\gamma}\cdot\bm g^{(2)}_{ab}-\bm \alpha\cdot\bm g^{(3)}_{ab}] \,,\nonumber\\
&V_a\bar V_b=\dfrac{1}{4}N_a N_b[\gamma^0f^{(0)}_{ab}+f^{(1)}_{ab}+\gamma^0\gamma^5f^{(2)}_{ab}-\gamma^5f^{(3)}_{ab}\nonumber\\
&+\gamma^0\bm
\Sigma\cdot\bm g^{(0)}_{ab}+\bm\Sigma\cdot\bm g^{(1)}_{ab}-\bm{\gamma}\cdot\bm g^{(2)}_{ab}+\bm \alpha\cdot\bm g^{(3)}_{ab}] \,,\nonumber\\
&V_a\bar U_b=\dfrac{1}{4}N_a N_b[\gamma^0\gamma^5f^{(0)}_{ab}+\gamma^5f^{(1)}_{ab}+\gamma^0f^{(2)}_{ab}-f^{(3)}_{ab}\nonumber\\
&-\bm \gamma\cdot\bm g^{(0)}_{ab}-\bm\alpha\cdot\bm g^{(1)}_{ab}+\gamma^0\bm{\Sigma}\cdot\bm g^{(2)}_{ab}-\bm \Sigma\cdot\bm g^{(3)}_{ab}] \,,\nonumber\\
&U_a\bar V_b=\dfrac{1}{4}N_a N_b[\gamma^0\gamma^5f^{(0)}_{ab}-\gamma^5f^{(1)}_{ab}+\gamma^0f^{(2)}_{ab}+f^{(3)}_{ab}\nonumber\\
&-\bm \gamma\cdot\bm g^{(0)}_{ab}+\bm\alpha\cdot\bm g^{(1)}_{ab}+\gamma^0\bm{\Sigma}\cdot\bm g^{(2)}_{ab}+\bm \Sigma\cdot\bm g^{(3)}_{ab}] \,.
\end{align}
Here
\begin{eqnarray}\label{eq3}
 &f^{(0)}_{ab}=(\bm P_a\bm P_b+1)A_{ab}-i[\bm P_a\times\bm P_b]\cdot\bm B_{ab}\,,\quad
f^{(2)}_{ab}=(\bm P_a+\bm P_b)\cdot\bm  B_{ab}\,,\nonumber\\
&\bm g^{(0)}_{ab}=(\bm  B_{ab}\cdot\bm P_a)\bm P_b+(\bm B_{ab}\cdot\bm P_b)\bm P_a-(\bm P_a\cdot\bm P_b-1)\bm  B_{ab}+i[\bm P_a\times\bm P_b]A_{ab}\,,\nonumber\\
&\bm g^{(2)}_{ab}=(\bm P_a+\bm P_b)A_{ab}-i[\bm  B_{ab}\times(\bm P_a-\bm P_b)]\,,\nonumber\\
&f^{(1)}_{ab}=-f^{(0)}_{ab}(\bm P_b\rightarrow -\bm P_b)\,,\quad f^{(3)}_{ab}=-f^{(2)}_{ab}(\bm P_b\rightarrow -\bm P_b)
\,,\nonumber\\
&\bm g^{(1)}_{ab}=-\bm  g^{(0)}_{ab}(\bm P_b\rightarrow -\bm P_b)\,,\quad \bm g^{(3)}_{ab}=-\bm g^{(2)}_{ab}(\bm P_b\rightarrow -\bm P_b)\,.
\end{eqnarray}
All  Feynman diagrams have a block structure and can be easily expressed via several quantities.
For one-photon emission (or absorption) and annihilation  these quantities are 
\begin{eqnarray}\label{eq4}
&\bar U_b\bm{\gamma}V_a=\bar V_b\bm{\gamma}U_a=N_aN_b\,\bm g^{(0)}_{ab}\,,\quad
&\quad\bar U_b\bm{\gamma}U_a=\bar V_b\bm{\gamma}V_a=N_aN_b\,\bm g^{(2)}_{ab}\,.
\end{eqnarray}
For two-photon emission (or absorption) and annihilation, the block terms are
\begin{eqnarray}\label{eq5}
&S^{(1)ij}_{ab}(k)\equiv(k^2+2kp_b)^{-1}\bar U_b\gamma^j(\hat p_b+\hat k+m)\gamma^iU_a=N_aN_b(k^2+2kp_b)^{-1}\nonumber\\
&\times\Big[2p_b^j g^{(2)i}_{ab}+(\delta^{ij}f^{(0)}_{ab}-i\epsilon^{ijl}g^{(0)l}_{ab})\,\omega
+i\epsilon^{ijl}k^lf^{(2)}_{ab}+g^{(2)j}_{ab}k^i+g^{(2)i}_{ab}k^j-\delta^{ij}\bm g^{(2)}_{ab}\cdot\bm k\Big]\,,\nonumber\\
&\nonumber\\
&S^{(2)ij}_{ab}(k)\equiv(k^2-2kp_b)^{-1}\bar V_b\gamma^j(-\hat p_b+\hat k+m)\gamma^iV_a=N_aN_b(k^2-2kp_b)^{-1}\nonumber\\
&\times\Big[-2p_b^j g^{(2)i}_{ab}+(\delta^{ij}f^{(0)}_{ab}-i\epsilon^{ijl}g^{(0)l}_{ab})\,\omega
+i\epsilon^{ijl}k^lf^{(2)}_{ab}+g^{(2)j}_{ab}k^i+g^{(2)i}_{ab}k^j-\delta^{ij}\bm g^{(2)}_{ab}\cdot\bm k\Big]\,,\nonumber\\
&\nonumber\\
&S^{(3)ij}_{ab}(k)\equiv(k^2+2kp_b)^{-1}\bar U_b\gamma^j(\hat p_b+\hat k+m)\gamma^iV_a=N_aN_b(k^2+2kp_b)^{-1}\nonumber\\
&\times\Big[2p_b^j g^{(0)i}_{ab}+(\delta^{ij}f^{(2)}_{ab}-i\epsilon^{ijl}g^{(2)l}_{ab})\,\omega
+i\epsilon^{ijl}k^lf^{(0)}_{ab}+g^{(0)j}_{ab}k^i+g^{(0)i}_{ab}k^j-\delta^{ij}\bm g^{(0)}_{ab}\cdot\bm k\Big]\,,\nonumber\\
&\nonumber\\
&S^{(4)ij}_{ab}(k)\equiv(k^2-2kp_b)^{-1}\bar V_b\gamma^j(-\hat p_b+\hat k+m)\gamma^iU_a=N_aN_b(k^2-2kp_b)^{-1}\nonumber\\
&\times\Big[-2p_b^j g^{(0)i}_{ab}+(\delta^{ij}f^{(2)}_{ab}-i\epsilon^{ijl}g^{(2)l}_{ab})\,\omega
+i\epsilon^{ijl}k^lf^{(0)}_{ab}+g^{(0)j}_{ab}k^i+g^{(0)i}_{ab}k^j-\delta^{ij}\bm g^{(0)}_{ab}\cdot\bm k\Big]\,.
\end{eqnarray}
By means of Eqs.~\eqref{eq1}-\eqref{eq5},    one can easily write  the explicit expressions for the amplitudes of huge amount of processes. These expressions are rather simple and have no any strong numerical cancellations. Therefore, they are very convenient for numerical evaluations.   

\begin{figure}
 \includegraphics[width=0.43\linewidth]{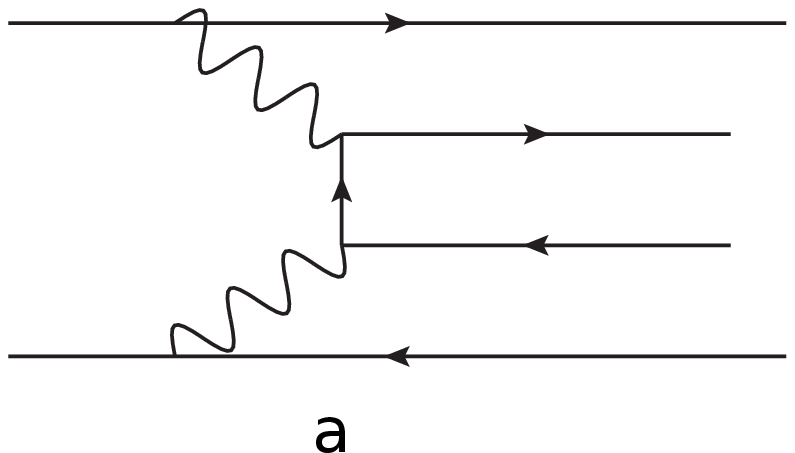}  
 \includegraphics[width=0.4\linewidth]{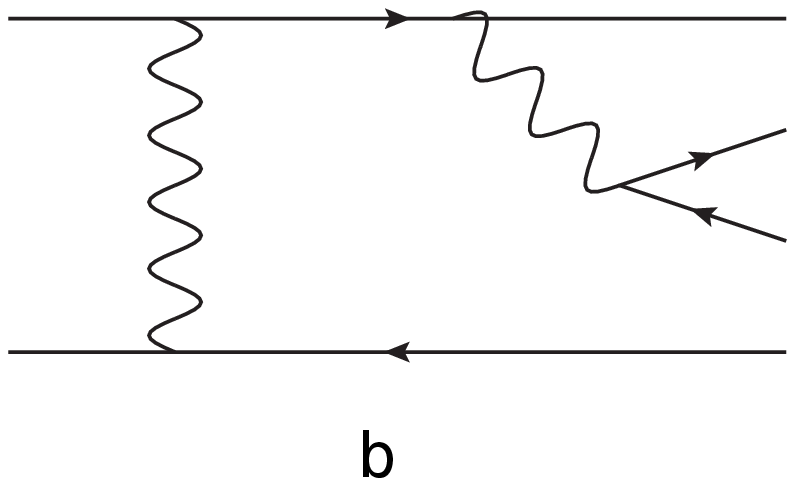} 
 \includegraphics[width=0.4\linewidth]{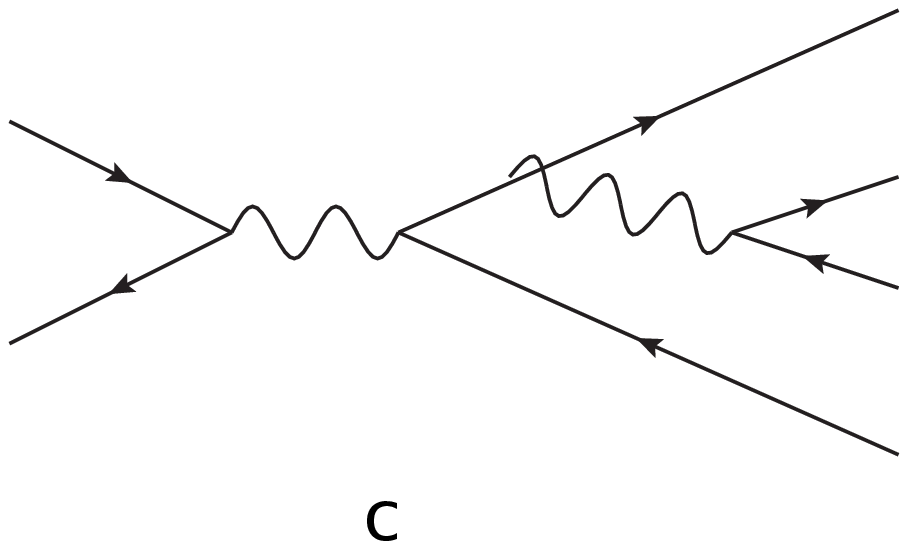}  
 \includegraphics[width=0.4\linewidth]{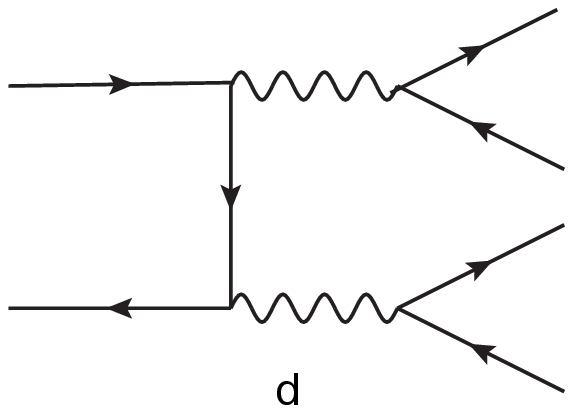}
	\caption{Typical diagrams   contributing to the amplitude of the process $e^+e^-\to e^+e^-e^+e^-$; wavy lines correspond to photons and straight lines to electrons and positrons. (a) Additional $e^+e^-$ pair is a result of annihilation of two virtual photons. (b)  Emission of  photon with its subsequent decay into $e^+e^-$ pair. (c)  One-photon annihilation of initial $e^+e^-$ pair with subsequent virtual photon  decay into two $e^+e^-$ pairs. (d)  Two-photon annihilation of initial $e^+e^-$ pair with subsequent decay of each virtual photon   into   $e^+e^-$ pair.}
	\label{diag} 
\end{figure}

Consider, as an example, the process $e^+e^-\to e^+e^-e^+e^-$.  Let  $\bm p_1,\,\bm p_2,\,\bm p_3 $ be the   momenta of initial electron and two final electrons, and  $\bm p_4,\,\bm p_5,\,\bm p_6 $ be the momenta of the initial positron and two final positrons, respectively. Typical Feynman diagrams, which contribute to the amplitude of the process, are shown in Fig.~\ref{diag}. 
The differential cross section, averaged over polarizations of  the initial particles and summed up   over polarizations of  the final  particles, reads
\begin{eqnarray}\label{cs}
&d\sigma= \dfrac{\alpha^4}{4\pi^4\,j}\,d\bm p_2 d\bm p_3 d\bm p_5 d\bm p_6\delta(p_2+p_3+p_5+p_6-p_1-p_4) \,
 \sum\limits_{\lambda_i}\,\left|\sum \limits_{a=1}^5     T^{(a)}_{\{\lambda_i\}} \right|^2\,,\nonumber\\
&T^{(a)}_{\{\lambda_i\}}=t^{(a)}_{123456}-t^{(a)}_{132456}-t^{(a)}_{123465}+t^{(a)}_{132465}\,,\quad j=\dfrac{\sqrt{(p_1p_4)^2-m^4}}{\varepsilon_1\varepsilon_4}\,.\label{ampl}
\end{eqnarray}
Here $\alpha$ is the fine structure constant, $t^{(1)}$ is the contribution  of the diagram shown in Fig.~\ref{diag}(a),  $t^{(2)}$ is   the contribution  of the diagram shown in Fig.~\ref{diag}(b), $t^{(3)}$ is similar to  $t^{(2)}$ but with the photon emission from positron line, $t^{(4)}$ corresponds to one-photon annihilation of initial $e^+e^-$ pair with subsequent virtual photon  decay into two $e^+e^-$ pairs  (Fig.~\ref{diag}(c)), and $t^{(5)}$ corresponds to two-photon annihilation of initial $e^+e^-$ pair with subsequent decay of each virtual photon   into   $e^+e^-$ pair (Fig.~\ref{diag}(d)).  

%

The amplitudes $t^{(a)}$ have the following form
\begin{eqnarray}\label{result}
&t^{(1)}_{123456}=\left(\prod\limits_{k = 1}^{6}N_k\right) \Big[g^{(2)j}_{12}-\dfrac{(p_2^j-p_1^j)(\bm p_2-\bm p_1,\bm g^{(2)}_{12}) }{(\varepsilon_2-\varepsilon_1)^2}\Big]\Big[g^{(2)i}_{54}-\dfrac{(p_5^i-p_4^i)(\bm p_5-\bm p_4,\bm g^{(2)}_{54}) }{(\varepsilon_5-\varepsilon_4)^2}\Big]\nonumber\\
&\times\dfrac{1}{(p_2-p_1)^2(p_5-p_4)^2}\Big[S^{(3)ij}_{63}(p_2-p_1)+S^{(3)ji}_{63}(p_5-p_4)\Big]\,,\nonumber\\
&t^{(2)}_{123456}=\left(\prod\limits_{k = 1}^{6}N_k\right)\Big[g^{(0)j}_{63}-\dfrac{(p_3^j+p_6^j)(\bm p_3+\bm p_6,\bm g^{(0)}_{63}) }{(\varepsilon_3+\varepsilon_6)^2}\Big]\Big[g^{(2)i}_{54}-\dfrac{(p_5^i-p_4^i)(\bm p_5-\bm p_4,\bm g^{(2)}_{54}) }{(\varepsilon_5-\varepsilon_4)^2}\Big]\nonumber\\
&\times\dfrac{1}{(p_3+p_6)^2(p_5-p_4)^2}\Big[S^{(1)ij}_{12}(p_3+p_6)+S^{(1)ji}_{12}(p_5-p_4)\Big]\,,\nonumber\\
&t^{(3)}_{123456}=\left(\prod\limits_{k = 1}^{6}N_k\right)\Big[g^{(0)j}_{63}-\dfrac{(p_3^j+p_6^j)(\bm p_3+\bm p_6,\bm g^{(0)}_{63}) }{(\varepsilon_3+\varepsilon_6)^2}\Big]\Big[g^{(2)i}_{12}-\dfrac{(p_2^i-p_1^i)(\bm p_2-\bm p_1,\bm g^{(2)}_{12}) }{(\varepsilon_2-\varepsilon_1)^2}\Big]\nonumber\\
&\times\dfrac{1}{(p_2-p_1)^2(p_3+p_6)^2}\Big[S^{(2)ij}_{54}(p_3+p_6)+S^{(2)ji}_{54}(p_2-p_1)\Big]\,,\nonumber\\
&t^{(4)}_{123456}=-\left(\prod\limits_{k = 1}^{6}N_k\right)\Big[g^{(0)j}_{63}-\dfrac{(p_3^j+p_6^j)(\bm p_3+\bm p_6,\bm g^{(0)}_{63}) }{(\varepsilon_3+\varepsilon_6)^2}\Big]\Big[g^{(0)i}_{14}-\dfrac{(p_1^i+p_4^i)(\bm p_1+\bm p_4,\bm g^{(0)}_{14}) }{(\varepsilon_1+\varepsilon_4)^2}\Big]\nonumber\\
&\times\dfrac{1}{(p_3+p_6)^2(p_1+p_4)^2}\Big[S^{(3)ij}_{52}(p_3+p_6)+S^{(3)ji}_{52}(-p_1-p_4)\Big]\,,\nonumber\\
&t^{(5)}_{123456}=-\left(\prod\limits_{k = 1}^{6}N_k\right)\Big[g^{(0)j}_{63}-\dfrac{(p_3^j+p_6^j)(\bm p_3+\bm p_6,\bm g^{(0)}_{63}) }{(\varepsilon_3+\varepsilon_6)^2}\Big]\Big[g^{(0)i}_{52}-\dfrac{(p_5^i+p_2^i)(\bm p_5+\bm p_2,\bm g^{(0)}_{52}) }{(\varepsilon_5+\varepsilon_2)^2}\Big]\nonumber\\
&\times\dfrac{1}{(p_3+p_6)^2(p_2+p_5)^2}\Big[S^{(4)ij}_{14}(p_3+p_6)+S^{(4)ji}_{14}(p_2+p_5)\Big]\,.
\end{eqnarray}

We now consider the process  $e^+e^-\rightarrow e^+e^-\mu^+\mu^-$.
 Let $\bm p_1$ and $\bm p_4$ are the initial electron and positron momenta,  
  $\bm p_2$ and $\bm p_5$ are momenta of the final electron and positron,  $\bm p_3$ and $\bm p_6$ are momenta of $\mu^-$ and $\mu^+$. The cross section of the processes $e^+e^-\rightarrow e^+e^-\mu^+\mu^-$ is given by Eq.~\eqref{ampl} with the substitution $T^{(a)}=t^{(a)}_{123456}$, where $t^{(a)}_{123456}$ is given by Eq.~\eqref{result} with $p_{1,2,4,5}^2=m_e^2$ and $p_{3,6}^2=m_\mu^2$.

\begin{figure}[h]
	\begin{minipage}[h]{0.45\linewidth}
		\center{\includegraphics[width=0.95\linewidth]{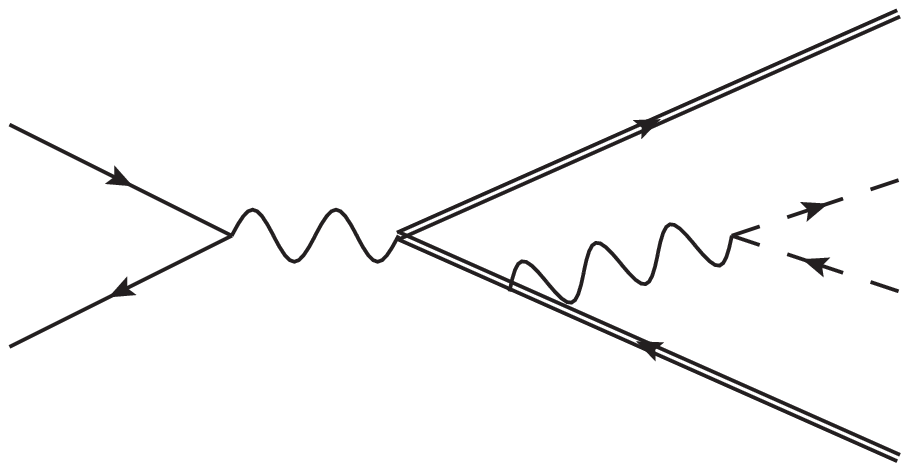}}  \\
	\end{minipage}
	\begin{minipage}[h]{0.45\linewidth}
		\center{\includegraphics[width=0.8\linewidth]{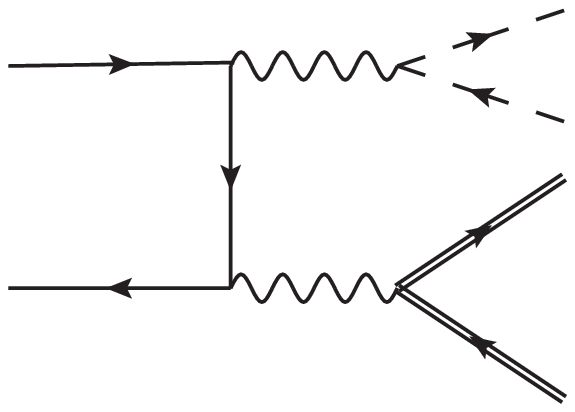}} \\
	\end{minipage}
	\caption{Two types of diagrams giving the main contribution to the amplitude of the process $e^+e^-\to \mu^+\mu^-\pi^+\pi^-$. Wavy lines correspond to photons, thin straight lines correspond to electrons and positrons, double lines correspond to muons, and dotted lines correspond to pions.}
	\label{diag1}
\end{figure}

It is easy to generalize our formulas to the case of polarized
particles, which is actual in context of the c-tau factory project~\cite{CTAU}.
Let electron and positron are described by the Dirac  spinors
$u_{a,\,\bm \zeta_a}$ and $v_{b,\,\bm \zeta_b}$,
where $\bm \zeta_a$ and $\bm \zeta_b$ are unit vectors directed along the electron and positron spin, respectively. Then we note that the following relations hold 
\begin{align} 
& \sum_{\lambda_a} \dfrac{\varepsilon_a}{m} U_{a, \lambda_a}\bar U_{a, \lambda_a}=\dfrac{\hat{p}_a+m}{2m}\,,\quad 
 \sum_{\lambda_b} \dfrac{\varepsilon_b}{m} V_{b, \lambda_b}\bar V_{b, \lambda_b}= \dfrac{\hat{p}_b-m}{2m}\,.
 \end{align}
Therefore, using the Dirac equation we can make the following transformation
 \begin{align} 
&u_{a,\,\bm \zeta_a}=\left(\sum_{\lambda_a} \dfrac{\varepsilon}{m} U_{a, \lambda_a}\bar U_{a, \lambda_a}\right)\cdot u_{a,\,\bm \zeta_a}=\sum_{\lambda_a} U_{a, \lambda_a}\,Z_{\lambda_a,\bm\zeta_a}\,, \nonumber\\
&{\bar v}_{b,\,\bm \zeta_b}=-{\bar v}_{b,\,\bm \zeta_b}\left(\sum_{\lambda_b} \dfrac{\varepsilon}{m} V_{b, \lambda_b}\bar V_{b, \lambda_b}\right) =\sum_{\lambda_b} {\bar V}_{b, \lambda_b}\,{\widetilde Z}_{\lambda_b,\bm\zeta_b} \,,\nonumber\\
  &Z_{\lambda_a,\bm\zeta_a}=\phi_{a\,,\lambda_a}^\dagger \phi_{a\,,\bm\zeta_a} \,, \quad
 {\widetilde Z}_{\lambda_b,\bm\zeta_b}=\chi_{b,\, \bm\zeta_b}^\dagger \phi_{b,\,\lambda_b} \,.
\end{align}
As a result, the cross section for the polarized initial electron and positron has the form
\begin{eqnarray}\label{cspol}
&d\sigma= \dfrac{\alpha^4}{ \pi^4\,j}\,d\bm p_2 d\bm p_3 d\bm p_5 d\bm p_6\delta(p_2+p_3+p_5+p_6-p_1-p_4) \,\nonumber\\
&\times \sum\limits_{\lambda_2,\lambda_3,\lambda_5,\lambda_6}\,\left| \sum\limits_{\lambda_1,\lambda_4 }\,\left( \sum \limits_{a=1}^5    T^{(a)}_{\{\lambda_i\}}\right) \,Z_{\lambda_1,\bm\zeta_1}{\widetilde Z}_{\lambda_4,\bm\zeta_4}\right|^2\,.
 \end{eqnarray}
Then we use the    relations
\begin{eqnarray}\label{rel}
&\phi_{1,\bm\zeta_1 } \phi_{1, \bm\zeta_1}^\dagger=\dfrac{1}{2}[1+(\bm\zeta_1\cdot\bm\sigma)]\,,\quad  \chi_{4,\bm\zeta_4 } \chi_{4, \bm\zeta_4}^\dagger=\dfrac{1}{2}[1-(\bm\zeta_4\cdot\bm\sigma)]\,,\nonumber\\
& \phi_{ \lambda' } \phi_{ \lambda}^\dagger=\dfrac{1}{2}(1+\lambda\sigma_3)\delta_{ \lambda' ,\lambda } +\dfrac{1}{2} (\bm \epsilon_{\lambda'}\cdot\bm\sigma)\delta_{ \lambda' ,-\lambda }\,,        \nonumber\\
&\bm \epsilon_{\lambda}=\bm e_x+i\lambda\bm e_y\,,
\end{eqnarray}
and the definition of $Z_{\lambda_a,\bm\zeta_a}$ and ${\widetilde Z}_{\lambda_b,\bm\zeta_b}$. We have
\begin{eqnarray}\label{ZZ}
&{\cal M}_{\lambda_1,\lambda_4,\lambda_1',\lambda_4'}= Z_{\lambda_1,\bm\zeta_1}{\widetilde Z}_{\lambda_4,\bm\zeta_4}  Z^*_{\lambda_1',\bm\zeta_1}{\widetilde Z}^*_{\lambda_4',\bm\zeta_4}=\dfrac{1}{4}\Big\{\delta_{ \lambda_1,\lambda_1'}\,\delta_{ \lambda_4,\lambda_4'}[1+\lambda_1(\bm\zeta_1\cdot\bm e_3)][1-\lambda_4(\bm\zeta_4\cdot\bm e_3)]\nonumber\\
&-\delta_{ \lambda_1,\lambda_1'}\,\delta_{ \lambda_4,-\lambda_4'}[1+\lambda_1\,(\bm\zeta_1\cdot\bm e_3)](\bm\zeta_4\cdot\bm \epsilon_{\lambda_4})+\delta_{ \lambda_1,-\lambda_1'}\,\delta_{ \lambda_4, \lambda_4'}(\bm\zeta_1\cdot\bm \epsilon_{\lambda_1'})[1-\lambda_4(\bm\zeta_4\cdot\bm e_3)]\nonumber\\
&-\delta_{ \lambda_1,-\lambda_1'}\,\delta_{ \lambda_4, -\lambda_4'}(\bm\zeta_1\cdot\bm \epsilon_{\lambda_1'})(\bm\zeta_4\cdot\bm \epsilon_{\lambda_4})\Big\}\,. 
\end{eqnarray}
Here $\bm \zeta_1$ and $\bm \zeta_4$ are the average polarization vectors in the electron and positron beams, respectively,  $|\bm \zeta_{1,4}|\le  1$. Finally we obtain  the explicit expression for the cross section with the polarized initial particles:
 \begin{align}\label{secpp}
 &d\sigma= \dfrac{\alpha^4}{ \pi^4\,j}\,d\bm p_2 d\bm p_3 d\bm p_5 d\bm p_6\delta(p_2+p_3+p_5+p_6-p_1-p_4) \,\nonumber\\
 &\times \sum\limits_{\lambda_i}   \sum\limits_{\lambda_i'}\,\left( \sum \limits_{a=1}^5    T^{(a)}_{\{\lambda_i\}}\right) \left( \sum \limits_{b=1}^5    T^{(b)*}_{\{\lambda_i' \}}\right)\delta_{\lambda_2,\lambda_2'} \delta_{\lambda_3,\lambda_3'} \delta_{\lambda_5,\lambda_5'}\delta_{\lambda_6,\lambda_6'}\, {\cal M}_{\lambda_1,\lambda_4,\lambda_1',\lambda_4'}\,.
\end{align}
 Another example is the   differential cross section of the process $e^+e^-\to \mu^+\mu^-\pi^+\pi^-$. Let $\bm p_1$ and $\bm p_4$ be the momenta $e^-$ and $e^+$, $\bm p_2$ and $\bm p_5$ are the momenta of  $\mu^-$ and $\mu^+$,  $\bm p_3$ and $\bm p_6$ are the momenta of $\pi^-$ and $\pi^+$. In this process, the main contribution to the amplitude is given by the two types of diagrams shown in Fig.~\ref{diag1}. Other contributions are suppressed by the pion electromagnetic form factor $F_{\pi}(q^2)$ with $q^2=(p_1+p_4)^2$. As a result, the differential cross section reads
\begin{align}
&d\sigma= \dfrac{\alpha^4}{16\pi^4\,j\varepsilon_3\varepsilon_6} \,d\bm p_2 d\bm p_3 d\bm p_5 d\bm p_6\delta(p_2+p_3+p_5+p_6-p_1-p_4)\, 
 \sum\limits_{\lambda_i=\pm 1}\left|t^{(4)}_{\pi}+t^{(5)}_{\pi}\right|^2\,,\nonumber\\
&t^{(4)}_{\pi}=-F_\pi\left((p_3+p_6)^2\right))(p_3^j+p_6^j)\Big[1-\dfrac{(\bm p_3+\bm p_6)^2}{(\varepsilon_3+\varepsilon_6)^2}\Big]\Big[g^{(0)i}_{14}-\dfrac{(p_1^i+p_4^i)(\bm p_1+\bm p_4,\bm g^{(0)}_{14}) }{(\varepsilon_1+\varepsilon_4)^2}\Big]\nonumber\\
&\times\dfrac{N_1N_2N_4N_5}{(p_3+p_6)^2(p_1+p_4)^2}\Big[S^{(3)ij}_{52}(p_3+p_6)+S^{(3)ji}_{52}(-p_1-p_4)\Big]\,,\nonumber\\
&t^{(5)}_{\pi}=-F_\pi\left((p_3+p_6)^2\right))(p_3^j+p_6^j)\Big[1-\dfrac{(\bm p_3+\bm p_6)^2}{(\varepsilon_3+\varepsilon_6)^2}\Big]\Big[g^{(0)i}_{52}-\dfrac{(p_5^i+p_2^i)(\bm p_5+\bm p_2,\bm g^{(0)}_{52}) }{(\varepsilon_5+\varepsilon_2)^2}\Big]\nonumber\\
&\times\dfrac{N_1N_2N_4N_5}{(p_3+p_6)^2(p_2+p_5)^2}\Big[S^{(4)ij}_{14}(p_3+p_6)+S^{(4)ji}_{14}(p_2+p_5)\Big]\,.
\end{align}
where a structure of the pion electromagnetic vertex has been taken into account.
Similarly, the cross section of the process 
$e^{+}e^{-} \to e^{+}e^{-}\pi^{+}\pi^{-}$ can easily be obtained. We do not present  the corresponding result here because of its bulkiness.
\section{Conclusion}
In the present work we have suggested a method to write the convenient representations for many  two-photon amplitudes. Our approach is based on the use of the gauge in which  the photon propagator $D_{\mu\nu}(k)$ has only space components ($D_{\mu 0}(k)=0$). The amplitudes obtained   have no any strong numerical cancellations and, therefore,  are very convenient for numerical evaluations. We have illustrated our approach on the examples of the processes  $e^+e^-\to e^+e^-e^+e^-$, $e^+e^-\rightarrow e^+e^-\mu^+\mu^-$, and $e^+e^-\to \mu^+\mu^-\pi^+\pi^-$.
It is shown how the results can be extended on the case of polarized particles. 

Basing on this approach the extension of BDK generator \cite{KMRS2019} was developed allowing for the simulation of  $e^+e^-\rightarrow 4l$    processes at increased numerical precision and of $e^+e^-\rightarrow l^+l^-\pi^+\pi^-$ processes with $\pi^+\pi^-$ pair in the $1^{--}$ state \cite{Shamov}. The contribution of narrow $1^{--}$ resonances are
accounted via the vacuum polarization effects.

\end{document}